\documentclass[conference]{IEEEtran}
\IEEEoverridecommandlockouts
\usepackage{cite}
\usepackage{amsmath,amssymb,amsfonts}
\usepackage{algorithmic}
\usepackage{graphicx}
\usepackage{textcomp}
\usepackage{xcolor}
\usepackage{multirow}
\usepackage{listings}
\usepackage{caption,subcaption}
\usepackage{booktabs}
\usepackage{enumitem}
\usepackage{graphicx}
\usepackage{tikz}
\usepackage{pifont}
\def\BibTeX{{\rm B\kern-.05em{\sc i\kern-.025em b}\kern-.08em
    T\kern-.1667em\lower.7ex\hbox{E}\kern-.125emX}}
\begin{document}

\title{Understanding Stress, Burnout, and Behavioral Patterns in Medical Residents Using Large-scale Longitudinal Wearable Recordings \\


}

\author{\IEEEauthorblockN{Tiantian Feng}
\IEEEauthorblockA{\textit{Department of Computer Science} \\
\textit{University of Southern California}\\
Los Angeles, CA \\
tiantiaf@usc.edu}
\and
\IEEEauthorblockN{Shrikanth Narayanan}
\IEEEauthorblockA{\textit{Department of Electrical Engineering} \\
\textit{University of Southern California}\\
Los Angeles, CA \\
shri@ee.usc.edu}
}

\maketitle

\begin{abstract}

Medical residency training is often associated with physically intense and emotionally demanding tasks, requiring them to engage in extended working hours providing complex clinical care. Residents are hence susceptible to negative psychological effects, including stress and anxiety, that can lead to decreased well-being, affecting them achieving desired training outcomes. Understanding the daily behavioral patterns of residents can guide the researchers to identify the source of stress in residency training, offering unique opportunities to improve residency programs. In this study, we investigate the workplace behavioral patterns of 43 medical residents across different stages of their training,  using longitudinal wearable recordings collected over a 3-week rotation. Specifically, we explore their ambulatory patterns, the computer access, and the interactions with mentors of residents. Our analysis reveals that residents showed distinct working behaviors in walking movement patterns and computer usage compared to different years in the program. Moreover, we identify that interaction patterns with mentoring doctors indicate stress, burnout, and job satisfaction.

\end{abstract}


\section{Introduction}

 \textit{Residency training} \cite{pratt2006constructing} serves as a fundamental component in medical education for training clinicians, aiming to provide medical students and other clinical trainees with practical experiences in specific specialties. Notably, residency learning is associated with intense demands that are cognitive, physical, emotional, and social while engaged in daily problem-solving and decision-making. The standard practice in residency training involves prolonged working routines (typically $>10$ hours) and complex working procedures. Beyond the extended working hours, residency learning is emotionally demanding, requiring residents to deliver empathy in patient care. Therefore, residents are susceptible to stress and anxiety, leading to declines in well-being, burnout symptoms, and negative training outcomes \cite{ishak2009burnout, prins2007burnout, raj2016well}. Increasingly, these negative impacts create challenges for residents in establishing the ability to make informed decisions in patient monitoring, medical treatment, and clinical rehabilitation.

Prior literature has revealed concerns about stress levels and burnout among residents. For example, researchers in \cite{kealy2016burnout} report that approximately $20\%$ of Canadian psychiatry residents face challenges caused by burnout, resulting in undesired training outcomes for residents. Moreover, a recent study \cite{hu2019discrimination} on American surgery residents demonstrates that mistreatment, such as verbal abuse, frequently occurs in resident training, leading to severe burnout and even triggering suicidal thoughts, particularly in female residents. However, existing studies on stress and burnout among residents rely on survey-based approaches, and there is a limited understanding of the behavioral patterns of residents and their associations with stress and burnout, preventing program administrators from identifying effective methods to improve residency training.

Recent technological advances in wearable sensors have enabled the emergence of applications for capturing diverse human behavioral patterns over a prolonged period in real-world settings \cite{saeb2015mobile, patel2012review, booth2019multimodal, wang2014studentlife, sano2018identifying, vaizman2018extrasensory}. In this study, we use wearable sensors to provide behavioral measurements of human movement, location proximity, and communicative interactions in residency training. We present results from computational data analysis from a three-week study of 43 medical residents in a hospital environment. We study the behaviors of residents in different program years through self-assessments and wearable sensor measurements, and we use the sensor data to identify factors associated with stress, anxiety, and burnout. In summary, our contributions include:

\begin{itemize}[leftmargin=*]
    \item One of the first-of-its-kind studies of stress and burnout with behavioral patterns among residents using wearables.

    \vspace{0.1mm}

    \item Comprehensive analysis of four important behavioral patterns: walking movement, human activity, computer accesses, and (mentor/supervisor) interactions.
    
    \vspace{0.1mm}
    
    \item Extended comparisons between residents at various training stages (first, second, and third years), revealing unique behavioral patterns with different experiences in the program.

\end{itemize}

\section{Study Background and Logistics}
\label{sec:study_design}

In late 2019 and early 2020, we conducted comprehensive experiments called TILES-2019 in a hospital workplace to examine the physiological, environmental, and behavioral variables impacting the wellness of medical residents \cite{yau2022tiles} as an extension to our previous TILES-2018 study on investigating wellness of general hospital employees (e.g., nurses, physicians, etc.) \cite{mundnich2020tiles}. Over three weeks, we collected data through wearable sensors and surveys from medical residents working in the ICU of the Los Angeles County and University of Southern California Medical Center (LAC+USC). Each participant has provided written consent to participate in the study. All study procedures were conducted in accordance with the University of Southern California's Institutional Review Board (IRB) approval (study ID HS-19-00606). The details of the datasets and research protocols are described in \cite{yau2022tiles}.

\section{Self-report Assessments}
\label{sec:baseline_variables}

\subsection{Pre-study Survey}
At the beginning of the study, we instructed participants to complete a set of baseline evaluations during the enrollment session to assess demographic information and behavioral
variables. For this paper, We consider the analysis on stress, depression, anxiety, and burnout. We summarize the surveys and scoring scales used in this work in Table~\ref{tab:self_report}.

\begin{itemize}[leftmargin=*]
    \vspace{0.5mm}
    \item\textbf{Stress} We adopted the \emph{Perceived Stress Scale} (PSS) \cite{cohen1983global} to evaluate stress of residents in the past month. The PSS includes ten items, each rated on a 5-point Likert scale from 0-4. The final PSS score was obtained by averaging ten items, with higher scores indicating higher stress levels.

    \vspace{0.5mm}
    \item\textbf{Burnout} We utilized \emph{Oldenburg Burnout Inventory} (OBI) \cite{Demerouti.JOPH.2010} to assess burnout. This survey measures burnout from disengagement and exhaustion, each comprising 8 items on a 4-point scale. We obtained the final scores through averaging, with higher scores representing higher burnout.

    \vspace{0.5mm}
    \item\textbf{Anxiety} We evaluated anxiety through the State-Trait Anxiety Inventory (STAI) \cite{Spielberger.STAI.1983}, consisting of 20 items on a 4-point scale. The final STAI was scored by summing responses, with higher scores indicating greater anxiety.

\end{itemize}

\subsection{Ecological Momentary Assessments (EMAs)}
    
In addition to pre-study surveys, EMAs were administered twice daily throughout the study. The midday EMA is a one-item survey evaluating stress, and the \emph{end-of-day EMA} assesses stress and work behaviors. Given the scope of this study, we include EMAs related to stress and job satisfaction.

\begin{itemize}[leftmargin=*]
    \item\textbf{Midday EMA} asked residents about stress in the morning on a 7-scale, with higher scores indicating higher stress.
    \vspace{0.5mm}
    
    \item\textbf{End-of-day EMA} instructed participants to report their stress during the afternoon on a scale from not at all stressed (1) to a great deal of stress (7). Moreover, participants were asked about their job satisfaction on a 6-scale item, with higher scores indicating lower job satisfaction. 

\end{itemize}

\begin{figure*}
    \centering
    \includegraphics[width=0.98\linewidth]{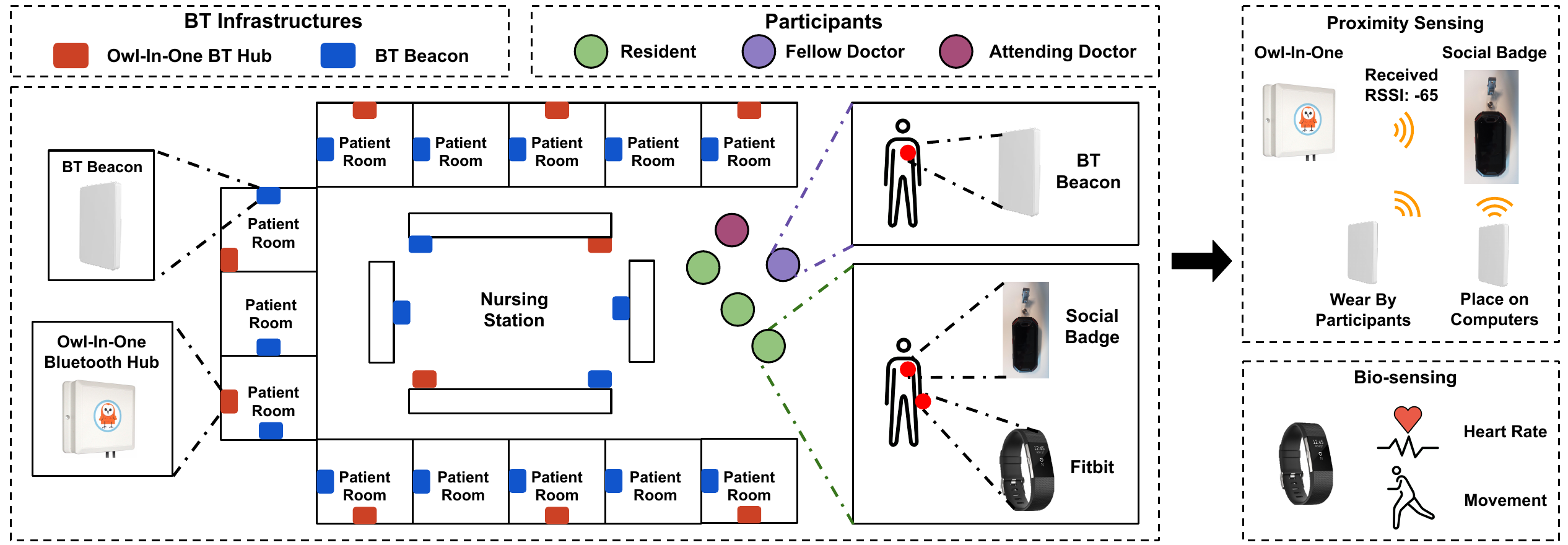}
    \caption{Sensor setup example in the hospital unit. Red and blue rectangular indicate the BT hubs and BT beacons, respectively. Each resident was instructed to wear a social badge and a Fitbit tracker, while the doctors were asked to carry BT beacons.}
    \vspace*{-0.95\baselineskip}
    \label{fig:layout}
\end{figure*}

\begin{table}[t]
    \caption{Self-report assessments used in this paper. $\uparrow$ indicates higher scales representing stronger behavioral variables.}
    \vspace{-1mm}
    \centering
    \footnotesize
    \begin{tabular*}{0.96\linewidth}{cccc}
        \toprule

        \multirow{1}{*}{\textbf{Survey Type}} & 
        \textbf{Behavioral Variables} & 
        \multirow{1}{*}{\textbf{Survey}} &
        \multirow{1}{*}{\textbf{Scales}} \\


        \midrule 

        \multirow{3}{*}{\textbf{Pre-study}} & Stress & \textbf{PSS\cite{cohen1983global}} & 0-4 ($\uparrow$) \\

        & Burnout & \textbf{OBI\cite{Demerouti.JOPH.2010}} & 1-4 ($\uparrow$) \\

        & Anxiety & \textbf{STAI\cite{Spielberger.STAI.1983}} & 20-80 ($\uparrow$) \\

        \cmidrule(lr){1-4}

        \textbf{Midday EMA} & Stress & - & 1-7 ($\uparrow$) \\

        \cmidrule(lr){1-4}
        \multirow{2}{*}{\textbf{End-of-day EMA}} & Stress & - & 1-7 ($\uparrow$) \\

        & Job Satisfaction & - & 1-6 ($\downarrow$) \\

        \bottomrule
    \end{tabular*}
    \vspace{-3.5mm}
    \label{tab:self_report}
\end{table}

\section{Wearable Sensing Devices}

We deployed two wearable sensors in this study. The first is a commercial wristband device (Fitbit Charge 3 \cite{fitbit}), while the other is a customized social badge sensor to track a participant's proximity along with Bluetooth (BLT) hubs and beacons installed throughout the hospital units. The complete data recording setup is shown in Figure~\ref{fig:layout}.

\vspace{0.5mm}
\noindent \textbf{Fitbit Charge $3$} measures physical activities, sleep quality, and heart rate through photoplethysmography (PPG). In this paper, we provide the analysis related to the heart rate (HR) and step count, which provide insights into behavioral patterns of physical activity and walking movement.

\vspace{0.5mm}
\noindent \textbf{The TILES Social Badge}
Participants wore a wearable badge-like sensor called TAR (TILES Audio Recorder) \cite{feng2018tiles} during each work shift throughout the study. TAR runs on a lightweight, budget-friendly Android platform called the Atom \cite{atom}. Our TAR application captures participant proximity to different target locations within each hospital unit. Specifically, the social badge is programmed to receive BLT packets from BLT beacons and hubs every 30 seconds, and the Received Signal Strength Indicator (RSSI) is used to determine the proximity data. The details about the sensor installation and proximity sensing are shown in Figure~\ref{fig:layout}. Specifically, stationary BLT beacons are placed on desktop and mobile computers, allowing us to estimate computer use. Lastly, we also recruited participants from the residents' supervisory/mentoring teams -- clinical fellows and attending doctors -- to wear BLT beacons while in the hospital units, enabling us to investigate the interaction between the supervisory doctors and residents. 

\begin{figure}
    \centering
    \includegraphics[width=0.85\linewidth]{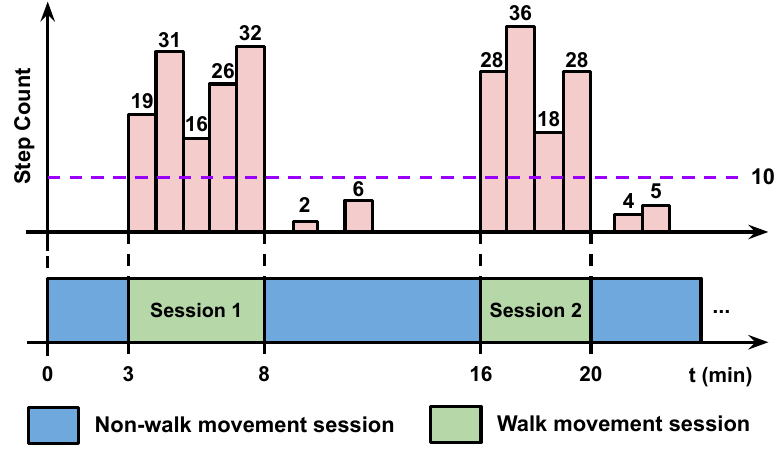}
    \caption{Example of the walking movement sessions.}
    \vspace{-3.5mm}
    \label{fig:walking}
\end{figure}

\section{Behavioral Feature Extraction}
\label{sec:method}

Our behavioral features are inspired by our previous works exploring daily activity and indoor proximity related to nurses \cite{feng2021multimodal, booth2019toward, feng2022exploring}. Specifically, we retain wearable recordings from 6 AM - 6 PM, and participants with data corresponding to less than $3$ shifts are excluded from further analysis.

\subsection{Movement and Activity}

\noindent \textbf{Walking movement} features were extracted by representing the walking movements as a time series of sessions following our past work in \cite{feng2022exploring}. We form a walking session as the time elapsed between the start and the end of a continuous walking movement. The minute-level step count data in a continuous walking movement are included if greater than $10$ steps due to noise that is common in wearable recordings. The recordings of the work shift with $10\%$ or more of missing step data are discarded for the remaining analysis. For example, Figure \ref{fig:walking} presents an example of walking movement sessions, with walking activity occurring at ${t}=\{3, 4, 5, 6, 7, 16, 17, 18, 19\}$. This leads to two walking sessions at ${t_{1}}=\{3, 4, 5, 6, 7\}$ and ${t_{2}}=\{16, 17, 18, 19\}$. We propose to use \textbf{session frequency} $f_{walk}$ (the number of unique sessions) and \textbf{total duration} $t_{walk}$ (the sum length of sessions) to measure the intensity of the walking movement. For instance, in the above example, the session frequency $f_{walk}=2$ and total duration $t_{walk}=9$min.

\vspace{0.5mm}
\noindent \textbf{Rest Activity} Similar to the approach taken in \cite{feng2021multimodal}, our study centers on activities characterized by minute-level heart rate readings. Specifically, we concentrate on studying the rest-activity patterns of residents. Similar to \cite{feng2021multimodal}, the rest-activity is defined as the heart rate below 50\% of its maximum. In this work, we study the total duration of rest-activity of residents.

\subsection{Proximity Detection}

\noindent \textbf{Computer access} is a frequent activity that residents engage in training. We form computer accesses into sessions similar to feature extraction of walking movement. Specifically, we regard proximity of $<$1 meter to a desktop or mobile computer as computer access, and continuous computer access with no more than a 2-minute gap in between is a computer access session. We use \textbf{session frequency} $f_{comp}$ and \textbf{total session duration} $t_{comp}$ to describe patterns of computer access.

\vspace{0.5mm}
\noindent \textbf{Mentoring/supervisory doctor interaction} is critical in residency training, and in our discussion with experts in residency programs, most residents have experienced stressful communications with mentoring doctors. In this study, we assume an interaction session between supervisory doctors and residents is defined as their proximity distance is $<$3 meters and continuous interaction of no more than a 2-minute gap in between. We use \textbf{session frequency} $f_{int}$to describe patterns of interactions between residents and doctors.

\begin{figure}[t] {
    \centering
    
    \begin{tikzpicture}

        \node[draw=none,fill=none] at (0,0){\includegraphics[width=0.45\linewidth]{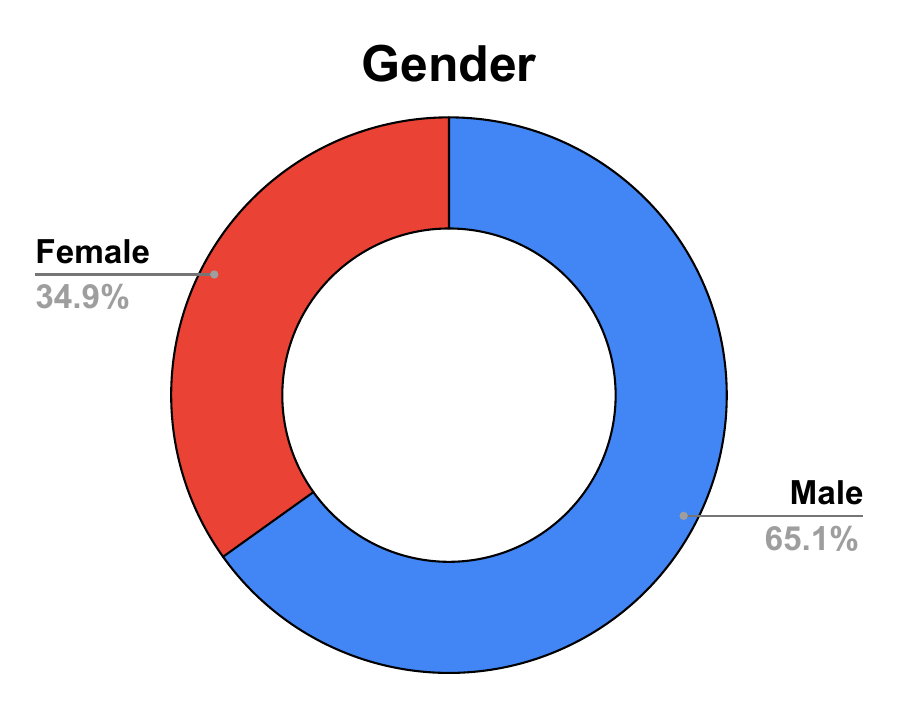}};

        \node[draw=none,fill=none] at (0.5\linewidth,0){\includegraphics[width=0.45\linewidth]{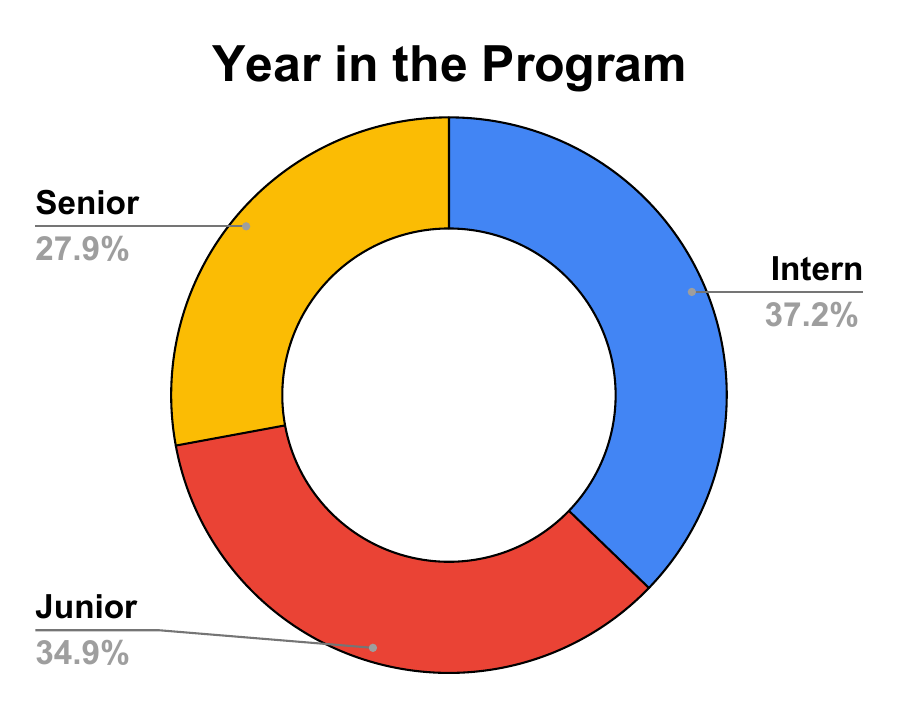}};

    \end{tikzpicture}
    \vspace{-4mm}
    \caption{Gender and year in the residents' program in this work.}
    \label{fig:demographics}
    \vspace{-3mm}
} \end{figure}

\section{Baseline Results}

\subsection{Demographics}

The analysis in this work included $n=43$ residents with a minimum of 3 days of Fitbit recordings, of which $n=28$ ($65\%$) were male, and $n=15$ ($35\%$) were female, as shown in Figure~\ref{fig:demographics}. Most participants enrolled in the Internal Medicine program ($n=36$, $84\%$), and the remaining residents were in the Emergency Medicine program ($n=7$, $16\%$). The year (stage) in the program ranges from $1$ to $3$, with $n=16$ participants as resident interns (first year), $n=15$ participants as junior residents, and $n=12$ participants as senior residents.

\subsection{Self-report Assessments}

\begin{figure}[t] {
    \centering
    
    \begin{tikzpicture}

        \node[draw=none,fill=none] at (0,2.5){\includegraphics[width=\linewidth]{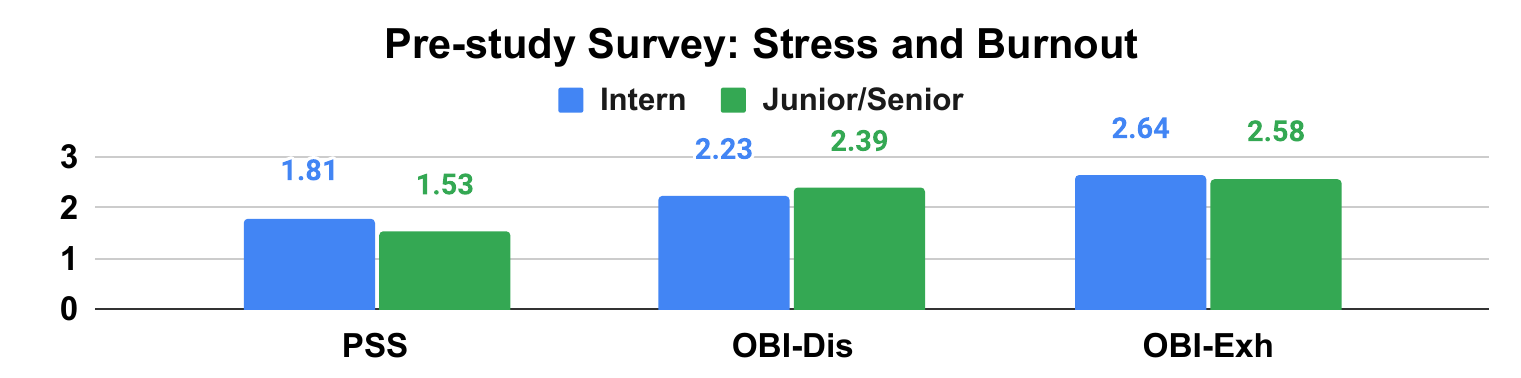}};

        \node[draw=none,fill=none] at (-2.1,0){\includegraphics[width=0.5\linewidth]{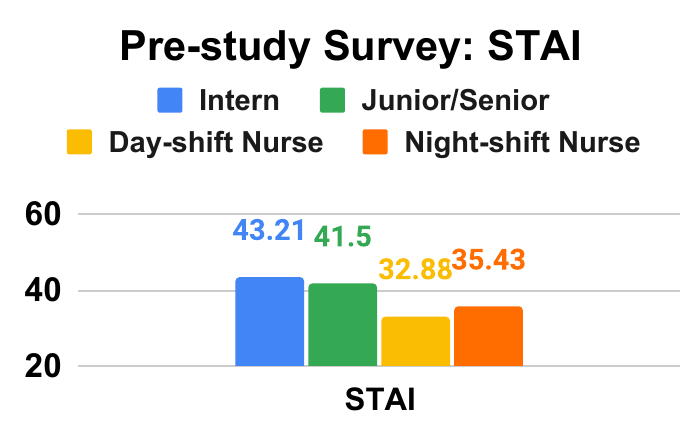}};

        \node[draw=none,fill=none] at (2.1,0){\includegraphics[width=0.5\linewidth]{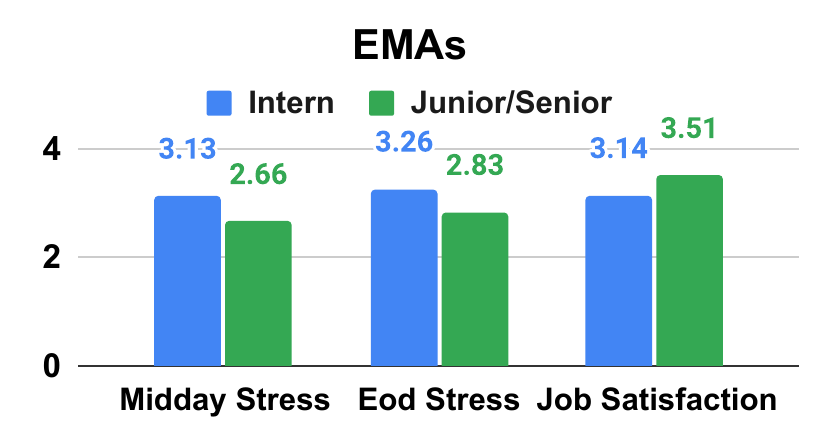}};

    \end{tikzpicture}
    \vspace{-6mm}
    \caption{Comparisons of pres-study assessments and EMAs.}
    \label{fig:surveys}
    \vspace{-3.5mm}
} \end{figure}

\noindent \textbf{Pre-study Survey} We plot the average scales of pre-study evaluation to the year in the program (intern; junior/senior) as demonstrated in Figure~\ref{fig:surveys}. We perform two-sample independent t-test results on the self-report assessment variables conditioned on the year in the program. The t-test shows that residents do not differ in the PSS, OBI, and STAI scores to the year in the program. Moreover, as we used the same survey to evaluate the anxiety levels of nursing professionals in \cite{feng2021multimodal}, we compare the STAI scores of residents with day and night shift nurses, revealing that residents have significantly higher anxiety scores than nurses (Residents: $42.10\pm5.21$; Day-shift Nurse: $32.88\pm7.63$; Day-shift Nurse: $35.43\pm9.08$; $p<0.01$). This result implies that residents experience greater anxiety levels than other healthcare providers in our study cohort.

\vspace{0.5mm}
\noindent \textbf{EMAs} The t-test shows that first-year residents and juniors/seniors do not differ in midday stress, end-of-day (eod) stress, and job satisfaction scores. However, we find that interns have higher stress levels than juniors/seniors on average.

\begin{table}
    \centering
    \caption{Comparisons of the movement and activity on training experiences with gender as the confounding factor. Statistical significance is denoted as $\mathbf{p^{*}<0.05}$. The mean and the std of each feature are denoted as mean (std).}

    \begin{tabular}{p{1.6cm}p{1.25cm}p{1.25cm}p{1.2cm}}

        \toprule
        \rule{0pt}{2ex}
        \multirow{1}{*}{Feature} & 
        \multicolumn{1}{c}{{Intern}} & 
        \multicolumn{1}{c}{{Junior/Senior}} &  
        \multicolumn{1}{c}{\multirow{1}{*}{{\centering p-value}}} \\ 
        \cmidrule(lr){1-1} \cmidrule(lr){2-3} \cmidrule(lr){4-4}
        
        \multicolumn{1}{l}{\textbf{Session Frequency}} & & & \rule{0pt}{2.25ex} \\
        
        \multicolumn{1}{l}{\hspace{0.25cm}{Walking Movement}} &
        \multicolumn{1}{c}{$62.8$ ($5.7$)} &
        \multicolumn{1}{c}{$68.5$ ($11.3$)} &
        \multicolumn{1}{c}{$\mathbf{<0.05^*}$} \rule{0pt}{2.25ex} \\
        
        \multicolumn{1}{l}{\textbf{Total Duration (min)}} & & & \rule{0pt}{2.25ex} \\

        \multicolumn{1}{l}{\hspace{0.25cm}{Walking Movement}} &
        \multicolumn{1}{c}{$148.1$ ($16.8$)} &
        \multicolumn{1}{c}{$161.9$ ($33.6$)} &
        \multicolumn{1}{c}{$0.112$} \rule{0pt}{2.25ex} \\

        \multicolumn{1}{l}{\hspace{0.25cm}{Rest-activity}} &
        \multicolumn{1}{c}{$582.5$ ($79.5$)} &
        \multicolumn{1}{c}{$625.8$ ($73.0$)} &
        \multicolumn{1}{c}{$0.075$}  \rule{0pt}{2.25ex} \\

        \bottomrule

    \end{tabular}
    \vspace{-1mm}
	\label{tab:activity}
\end{table}

\subsection{Shift-level Behavioral Patterns}

\noindent \textbf{Movement and Activity} Here, we perform two-way ANOVA to examine the effect on the behavioral variables in walking movement and rest-activity from the year in the program (intern; non-intern) with gender (male; female), as demonstrated in Table~\ref{tab:activity}. The comparisons demonstrate that junior and senior residents engaged in more walking sessions--$f_{walk}$--than intern residents (Intern: $62.8\pm5.7$; Junior/Senior: $68.5\pm11.3$; $p<0.05$). However, we identify that resident interns and juniors/seniors do not differ in $f_{rest}$, $t_{rest}$, and $t_{walk}$.

\vspace{0.5mm}
\noindent \textbf{Computer Access} A total of $37$ participants with more than 3 days of proximity recordings is available for the analysis presented in Table~\ref{tab:proximity}. The comparisons show no differences in computer access to training experiences. However, we observe that residents spend a substantial time accessing computers. 

\vspace{0.5mm}
\noindent \textbf{Mentoring Doctor Interaction}
The comparisons in Table~\ref{tab:proximity} show that senior residents engaged in more frequent interactions with attending doctors than first-year interns, while residents do not differ in fellow doctor interactions.

\begin{figure}
    \centering
    \vspace{-1mm}
    \includegraphics[width=\linewidth]{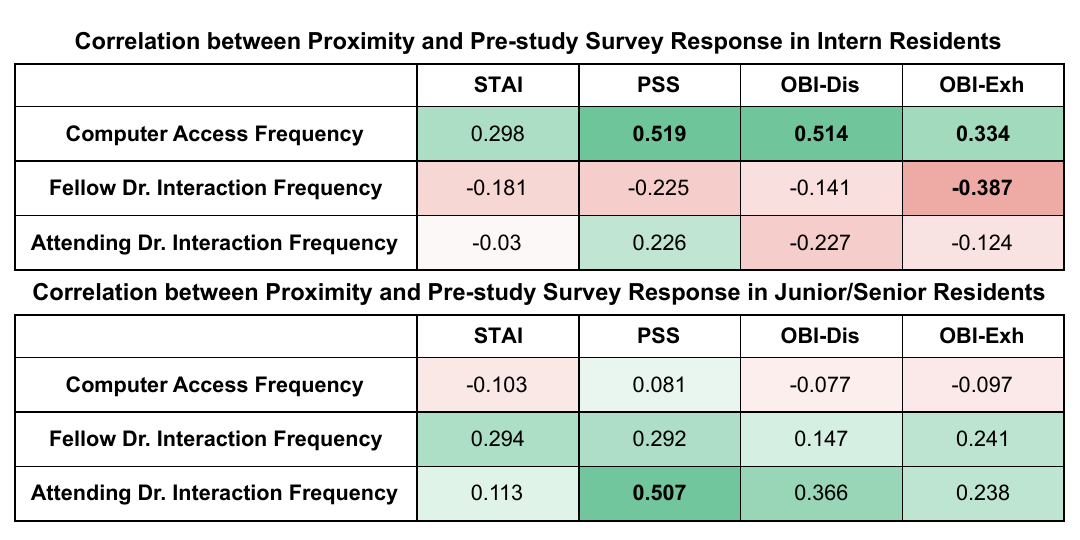}
    \caption{Correlations of computer access and mentoring doctor interaction with pre-study survey responses.}
    \vspace*{-0.95\baselineskip}
    \label{fig:corr}
\end{figure}

\vspace{0.5mm}
\noindent \textbf{Are computer use and doctor interactions correlated with pre-study surveys?} We plot the correlation between proximity-derived features and pre-study survey responses in Figure~\ref{fig:corr}. The results show that computer access positively correlates with stress and disengagement in intern residents. Moreover, we identify that the interaction frequency with attending doctors positively correlates with stress among junior and senior residents. Considering that junior and senior residents engaged in more interactions with attending doctors than intern residents, this implies that interaction with attending doctors may be an important indicator of stress.

\begin{table}
    \centering
    \caption{Comparisons of the proximity information on training experiences. Statistical significance is denoted as $\mathbf{p^{*}<0.05}$. We denote the mean and the std as mean (std).}

    \begin{tabular}{p{1.6cm}p{1cm}p{1cm}p{1.2cm}}

        \toprule
        \rule{0pt}{2ex}
        \multirow{1}{*}{Feature} & 
        \multicolumn{1}{c}{{Intern}} & 
        \multicolumn{1}{c}{{Junior/Senior}} &  
        \multicolumn{1}{c}{\multirow{1}{*}{{\centering p-value}}} \\ 
        \cmidrule(lr){1-1} \cmidrule(lr){2-3} \cmidrule(lr){4-4}
        
        \multicolumn{1}{l}{\textbf{Session Frequency}} & & & \rule{0pt}{2.25ex} \\
        
        \multicolumn{1}{l}{\hspace{0.15cm}{Computer Access}} &
        \multicolumn{1}{c}{$56.1$ ($8.4$)} &
        \multicolumn{1}{c}{$60.4$ ($7.3$)} &
        \multicolumn{1}{c}{$0.121$} \rule{0pt}{2.25ex} \\
        
        \multicolumn{1}{l}{\hspace{0.15cm}{Fellow Dr. Interaction}} &
        \multicolumn{1}{c}{$27.1$ ($7.4$)} &
        \multicolumn{1}{c}{$30.6$ ($6.2$)} &
        \multicolumn{1}{c}{0.139}  \rule{0pt}{2.25ex} \\

        \multicolumn{1}{l}{\hspace{0.15cm}{Attending Dr. Interaction}} &
        \multicolumn{1}{c}{$16.2$ ($3.0$)} &
        \multicolumn{1}{c}{$20.2$ ($5.0$)} &
        {$\mathbf{<0.05^*}$} \rule{0pt}{2.25ex} \\

        \multicolumn{1}{l}{\textbf{Total Duration (min)}} & & & \rule{0pt}{2.25ex} \\

        \multicolumn{1}{l}{\hspace{0.15cm}{Computer Access}} &
        \multicolumn{1}{c}{$279.5$ ($76.9$)} &
        \multicolumn{1}{c}{$312.1$ ($67.0$)} &
        \multicolumn{1}{c}{$0.196$} \rule{0pt}{2.25ex} \\

        \bottomrule

    \end{tabular}
    \vspace{-3.5mm}
	\label{tab:proximity}
\end{table}

\begin{table*}[t]
    \caption{A typical day of residents in residency training.}
    \vspace{-1mm}
    \centering
    \footnotesize
    \begin{tabular*}{\linewidth}{ccc}
        \toprule
        \textbf{Workflows} & 
        \textbf{Time in a Day} &
        \textbf{Responsibilities and Actions} \\

        \midrule 
        \multirow{1}{*}{\textbf{Morning Reviews}} & \multirow{1}{*}{$6$am-$6$:$30$am} & Review sign-out forms from night-shift staff \\

        \cmidrule(lr){1-3}

        \multirow{1}{*}{\textbf{Pre-round}} & \multirow{1}{*}{$6$:$30$am-$8$:$30$am} & Review lab results / Talk with nursing staff / Visit patients to perform brief assessments \\

        \cmidrule(lr){1-3}
        
        \multirow{1}{*}{\textbf{Morning Rounds}} & \multirow{1}{*}{$8$:$30$am-$12$pm} & Team up with peer residents and mentoring doctors / Review updates from patients / Mentoring doctors teach \\ 

        \cmidrule(lr){1-3}
        \multirow{1}{*}{\textbf{Lunch Break}} & \multirow{1}{*}{$12$pm-$1$pm} & Have lunch / Take Break / Additional didactic learning session \\

        \cmidrule(lr){1-3}
        \multirow{1}{*}{\textbf{Afternoon Notes}} & \multirow{1}{*}{$1$pm-$3$pm} & Write notes and review updated lab results \\

        \cmidrule(lr){1-3}
        \multirow{1}{*}{\textbf{Afternoon Rounds}} & \multirow{1}{*}{$3$pm-$4$pm} & Team up with fellow and attending doctors / Review updates from patients / Attending and fellow doctors teach \\

        \cmidrule(lr){1-3}
        \multirow{1}{*}{\textbf{Shift Transition}} & \multirow{1}{*}{$4$pm-$6$pm} & Finish notes /  Provide sign-out forms for night-shift staff \\

        \bottomrule
    \end{tabular*}
    \vspace{-4mm}
    \label{tab:residency_training}
\end{table*}

\section{Nuanced Behavioral Patterns}

\subsection{A Day in the Life of Residents during Residency Training}
\label{sec:mapping}

To better understand the behaviors of residents, we consulted with experts who have experience in residency training to provide us with knowledge about typical workflows for residents in daily training, as reported in Table~\ref{tab:residency_training}. It is worth noting that different hospitals may have minor workflow differences compared to those presented in this paper. 

\vspace{0.5mm}
\begin{itemize}[leftmargin=*]
    \item \textbf{Morning workflows} A resident typically arrives at the hospital at 6 am. Upon arrival at the hospital, the resident reviews the sign-out forms about critical events from each patient overnight for about 30 minutes. Then, the resident engaged in a 2-hour pre-round involving reviewing lab results, talking with nurses, and performing brief patient assessments. Starting at 8:30 am, residents participate in morning rounds with resident colleagues, fellow doctors, and attending doctors as a team. The starting time of the round depends on the attending doctor's schedule. The team visits each patient during morning rounds to discuss overnight events and lab results. The length of the morning round is based on the complexity and the number of patients. 
    
    \vspace{0.5mm}

    \item \textbf{Lunch Breaks} During lunch hour (between 12 pm and 1 pm), residents have lunch but may also engage in didactic teaching sessions such as case conferences.
    
    \vspace{0.5mm}
    
    \item \textbf{Afternoon workflows} After lunch, residents continue with notes for 2 hours. At 3 pm, residents and mentoring doctors perform afternoon rounds that review updates on each patient. This session is not as lengthy as morning rounds and typically ends in one hour. At the end of the shift, the residents provide sign-out forms for overnight staff.

\end{itemize}

\begin{figure}[t] {
    \centering
    
    \begin{tikzpicture}

        \node[draw=none,fill=none] at (0,0){\includegraphics[width=0.52\linewidth]{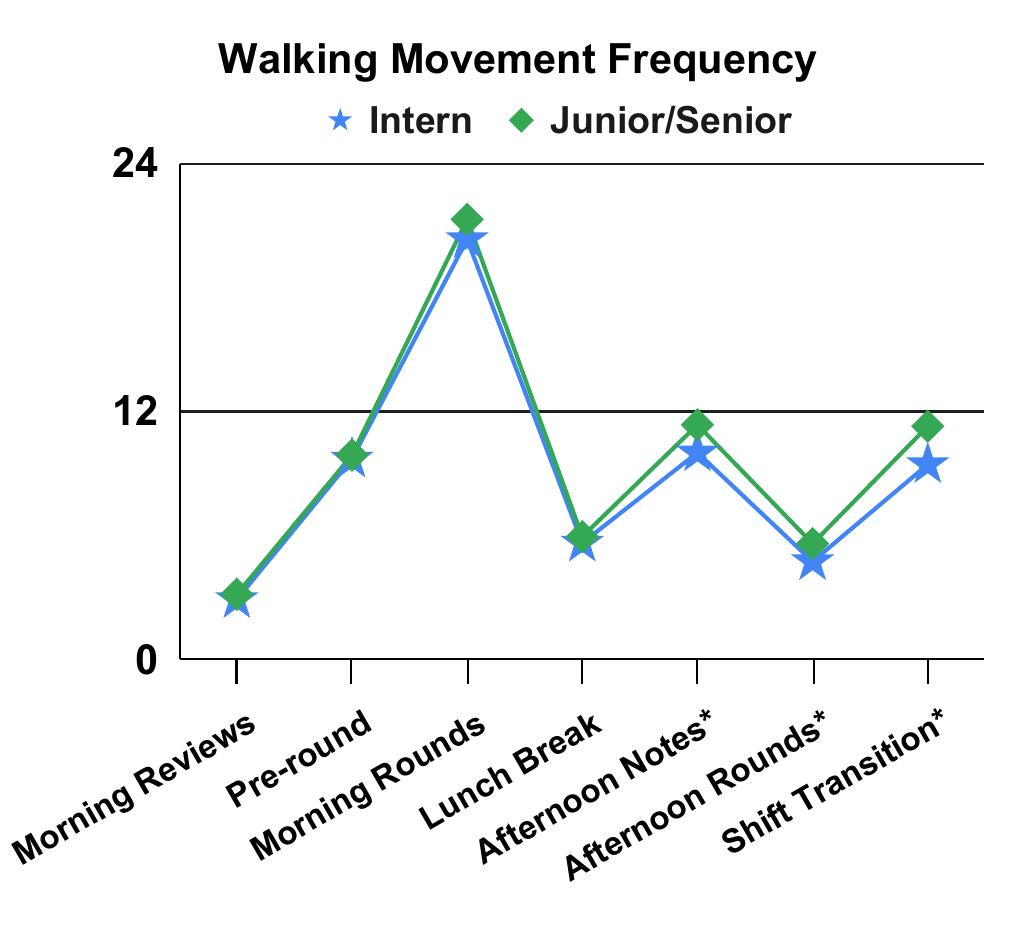}};

        \node[draw=none,fill=none] at (0.48\linewidth,0){\includegraphics[width=0.52\linewidth]{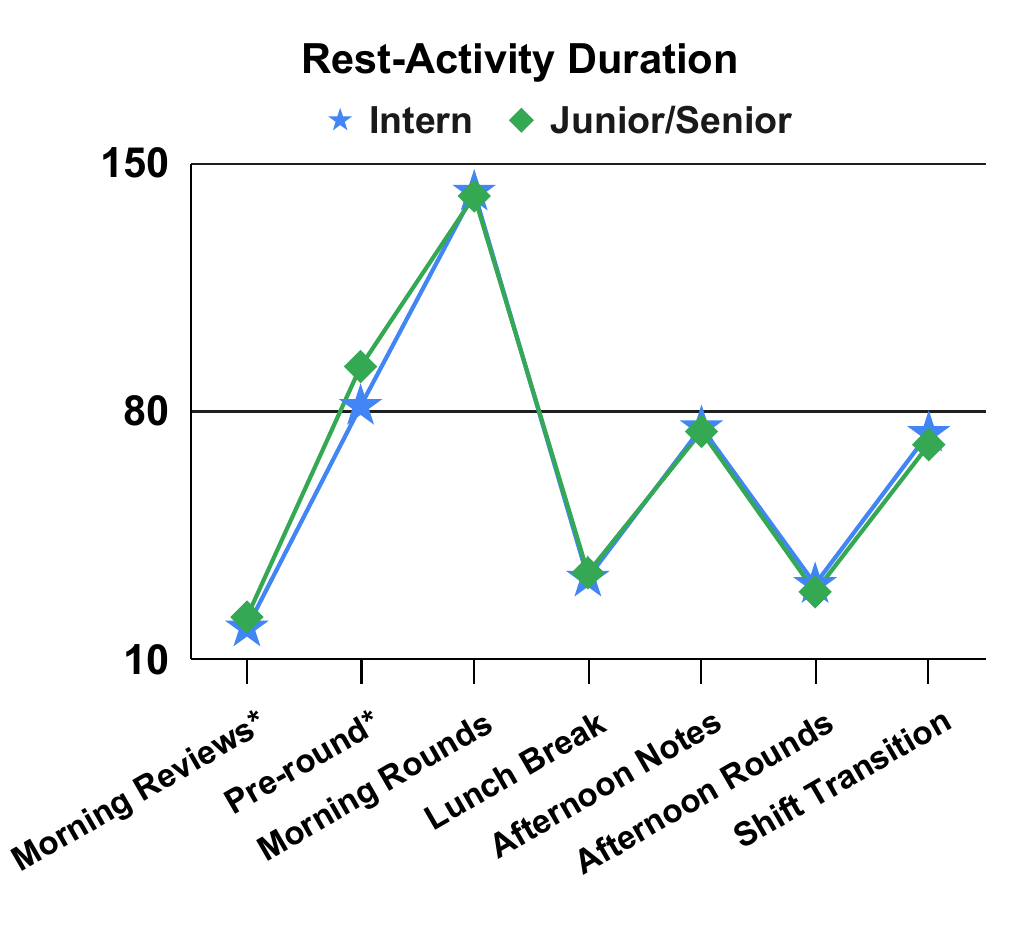}};

    \end{tikzpicture}
    \vspace{-5mm}
    \caption{Comparisons of walk and rest-activity at workflow level, with $\mathbf{^*}$ indicating statistical significance between resident interns and those in junior and senior years.}
    \label{fig:workflow_walk}
    \vspace{-3.5mm}
} \end{figure}

\begin{figure}[t] {
    \centering
    
    \begin{tikzpicture}

        \node[draw=none,fill=none] at (0,0){\includegraphics[width=\linewidth]{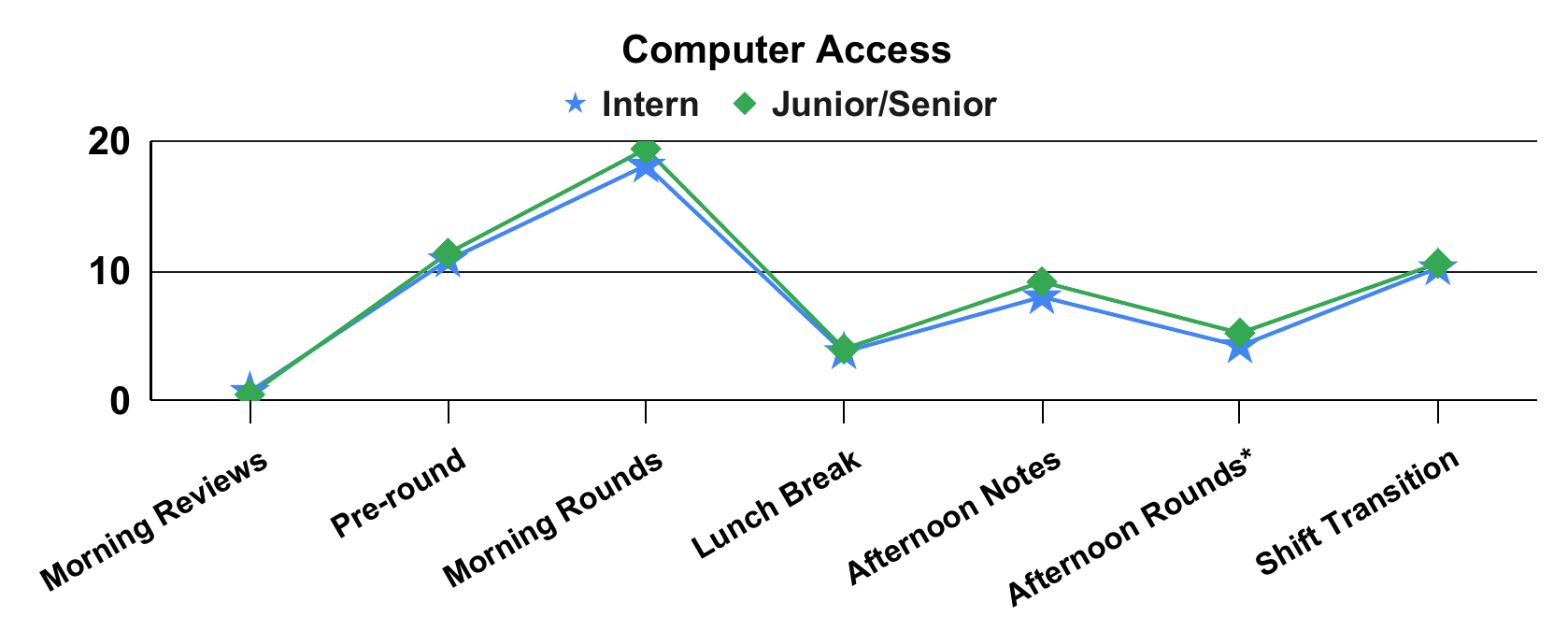}};

    \end{tikzpicture}
    \vspace{-5mm}
    \caption{Comparisons of computer access at workflow level between resident interns and those in junior and senior years. $\mathbf{^*}$ indicates statistical significance.}
    \label{fig:workflow_computer}
    \vspace{-3mm}
} \end{figure}

\begin{figure}
    \centering
    \vspace{-2.5mm}
    \includegraphics[width=\linewidth]{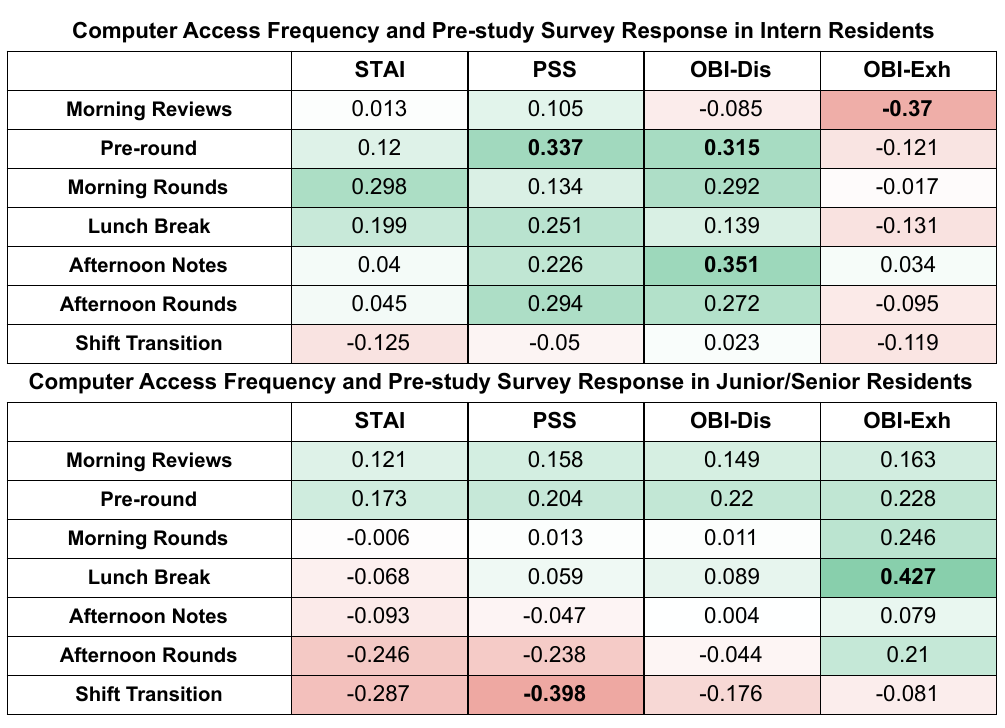}
    \caption{Correlations between computer access and self-report survey responses in different years in the program.}
    \vspace*{-0.95\baselineskip}
    \label{fig:corr_computer}
\end{figure}

\subsection{Workflow-level Behavioral Patterns}

\noindent \textbf{Movement and Activity} We plot the comparisons between resident interns and those in junior and senior years in Figure~\ref{fig:workflow_walk}. Similar to shift-level analysis, we conduct a two-way ANOVA to examine the effect on the movement and rest-activity from the year in the program with gender at the workflow level. The comparisons show that junior and senior residents engaged in more frequent walking movements in the afternoon, leading to overall higher $f_{walk}$ presented in Table~\ref{tab:activity}. Moreover, interns have less rest-activity in the morning reviews and pre-round than juniors and seniors.

\begin{figure}[t] {
    \centering
    
    \begin{tikzpicture}

        \node[draw=none,fill=none] at (0,0){\includegraphics[width=0.52\linewidth]{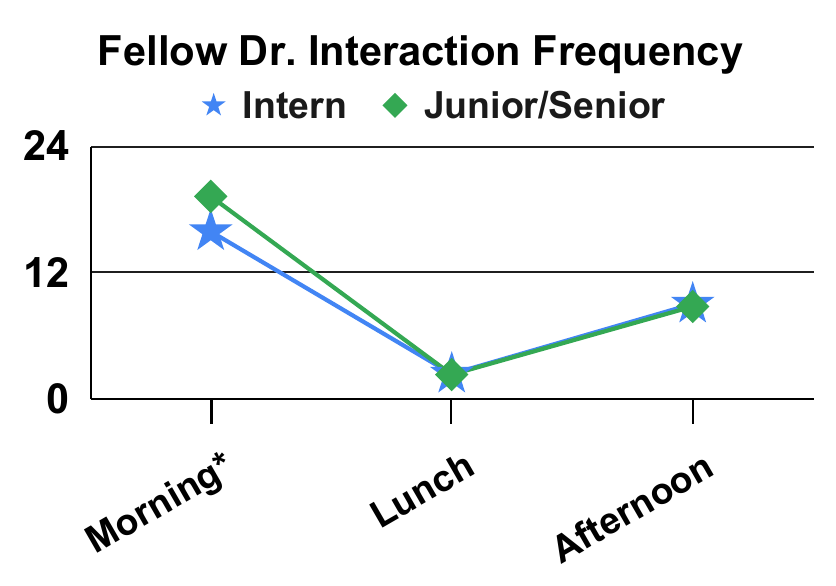}};

        \node[draw=none,fill=none] at (0.48\linewidth,0){\includegraphics[width=0.52\linewidth]{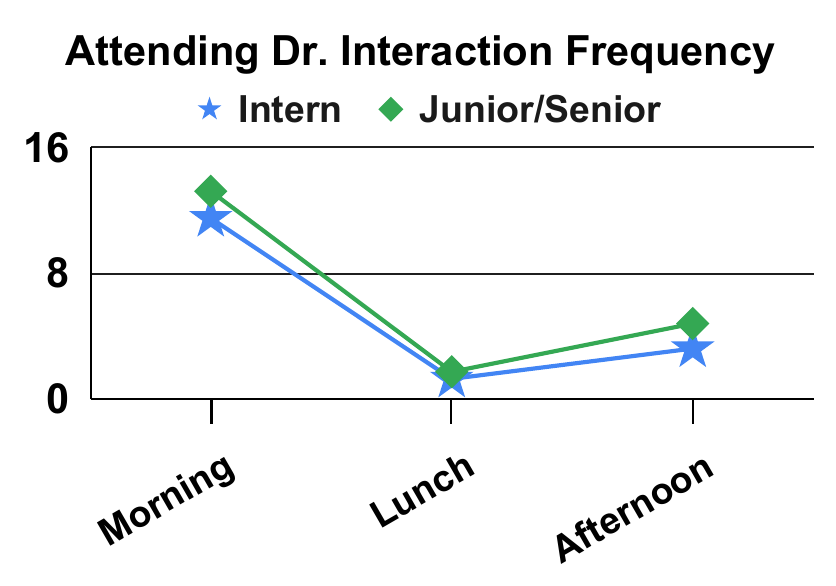}};

    \end{tikzpicture}
    \vspace{-5mm}
    \caption{Comparisons of doctor interaction patterns at workflow level. $\mathbf{^*}$ indicates statistical significance between resident interns and those in junior and senior years.}
    \label{fig:workflow_doctor}
    \vspace{-3mm}
} \end{figure}

\begin{figure}
    \centering
    \vspace{-2.5mm}
    \includegraphics[width=\linewidth]{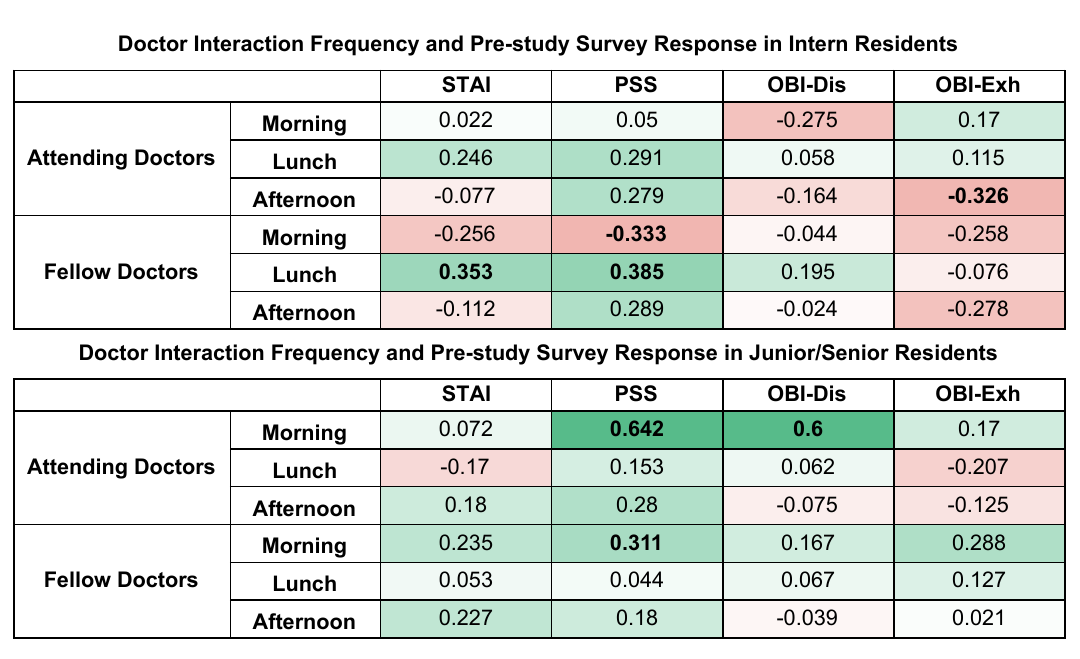}
    \caption{Correlations between mentoring doctor interaction and pre-study survey responses in different years in the program.}
    \vspace*{-0.95\baselineskip}
    \label{fig:corr_doctor}
\end{figure}

\vspace{0.5mm}
\noindent \textbf{Computer Access} Figure~\ref{fig:workflow_computer} compares the computer access with the year in the program. We observe that junior and senior residents have more frequent computer use than interns during the afternoon rounds (Intern: $4.21\pm1.27$; Junior/Senior: $5.21\pm1.14$; $p<0.05$). We further study the correlations between workflow computer access features and self-reported assessments in Figure~\ref{fig:corr_computer}. The correlation results indicate that intern residents tend to feel more stressed with frequent computer access in pre-round sessions. One plausible explanation is that interns are familiarizing themselves with training workflows, and more frequent computer use in the pre-round may decrease the time spent talking to nurses or performing patient assessments, interfering with discussions in the subsequent morning rounds. Moreover, among junior and senior residents, stress levels negatively correlate with computer use during the shift transition.

\vspace{0.5mm}
\noindent \textbf{Mentoring Doctor Interaction} Figure~\ref{fig:workflow_doctor} plots the interaction frequency between doctors and residents with the stage in the program. As the mentoring doctor interaction mainly occurs in certain workflows (e.g., morning rounds), we combine workflows before and after lunch as morning and afternoon, respectively. The comparisons indicate junior and senior residents interact more frequently with fellow doctors in the morning than interns. Moreover, we observe junior and senior residents interact more frequently with attending doctors throughout the day, but this difference is not statistically significant. Similar to computer access, we explore the correlation between interaction patterns and self-report assessment. The correlation results indicate that junior and senior residents tend to report higher stress levels with frequent interactions with mentoring doctors in the morning. One plausible reason might be related to morning rounds, where junior and senior residents lead more discussions with mentoring doctors.

\section{Machine Learning Validation}

Here, we use an ML experiment using behavioral features to predict the self-reported EMAs described in section~\ref{sec:baseline_variables}.

\subsection{Experiment Details}
For the modeling, we apply workflow-level features in walking movement, rest activity, computer access, and doctor interaction. We use the features before lunch to predict midday stress and features during lunch and after lunch to predict end-of-day stress and job satisfaction. We added the training stage and gender as features. On the other hand, we binarize the self-report variables using the median value of each measure to create a balanced label distribution. Specifically, we use the efficacy of the Random Forest (RF) classifier for the modeling. We impute the missing value of the feature with mean imputation and perform z-normalization on each feature to remove scaling differences. We apply a 3-fold cross-validation scheme combined with grid search to evaluate the performance and report the best results in macro-F1 scores.

\begin{table}
    \centering
    \caption{Table showing the prediction performance (macro-F1) of EMAs using Random Forest with behavioral features.}

    \begin{tabular}{p{2cm}p{1.2cm}p{1.2cm}p{1.2cm}}

        \toprule
        \multicolumn{1}{c}{Features} & 
        \multicolumn{1}{c}{Midday Stress} & 
        \multicolumn{1}{c}{End-of-day Stress} &  
        \multicolumn{1}{c}{Job Satisfaction} \\
        \cmidrule(lr){1-1} \cmidrule(lr){2-4} \cmidrule(lr){3-3} \cmidrule(lr){4-4}
        
        \multicolumn{1}{l}{{Walk Movement}} &
        \multicolumn{1}{c}{$52.55$} &
        \multicolumn{1}{c}{$54.70$} &
        \multicolumn{1}{c}{$57.10$}  \\

        \multicolumn{1}{l}{{Rest-activity}} &
        \multicolumn{1}{c}{$52.95$} &
        \multicolumn{1}{c}{$\mathbf{63.41}$} &
        \multicolumn{1}{c}{$54.15$}  \\

        \multicolumn{1}{l}{{Computer Use}} &
        \multicolumn{1}{c}{$\mathbf{56.20}$} &
        \multicolumn{1}{c}{$54.11$} &
        \multicolumn{1}{c}{$56.37$}  \\

        \multicolumn{1}{l}{Dr. Interaction} &
        \multicolumn{1}{c}{$49.58$} &
        \multicolumn{1}{c}{$45.59$} &
        \multicolumn{1}{c}{$54.77$}  \\

        \multicolumn{1}{l}{Multimodal} &
        \multicolumn{1}{c}{$52.81$} &
        \multicolumn{1}{c}{$57.70$} &
        \multicolumn{1}{c}{$\mathbf{60.89}$}  \\

        \bottomrule

    \end{tabular}
    
    \label{tab:ml_result}
    \vspace{-3mm}
\end{table}

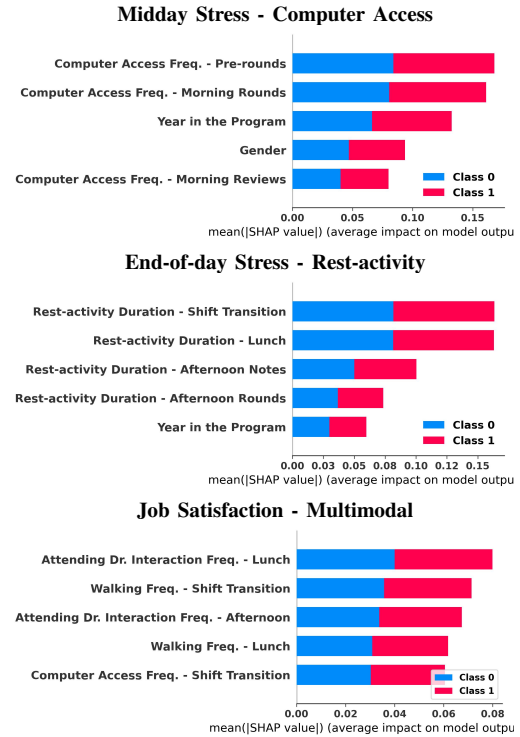
\begin{figure}[t] {
    \centering
    
    \begin{tikzpicture}

        \node at (0.2,8.2) {\footnotesize \textbf{Midday Stress - Computer Access}};
        \node[draw=none,fill=none] at (0,6.6){\includegraphics[width=0.76\linewidth]{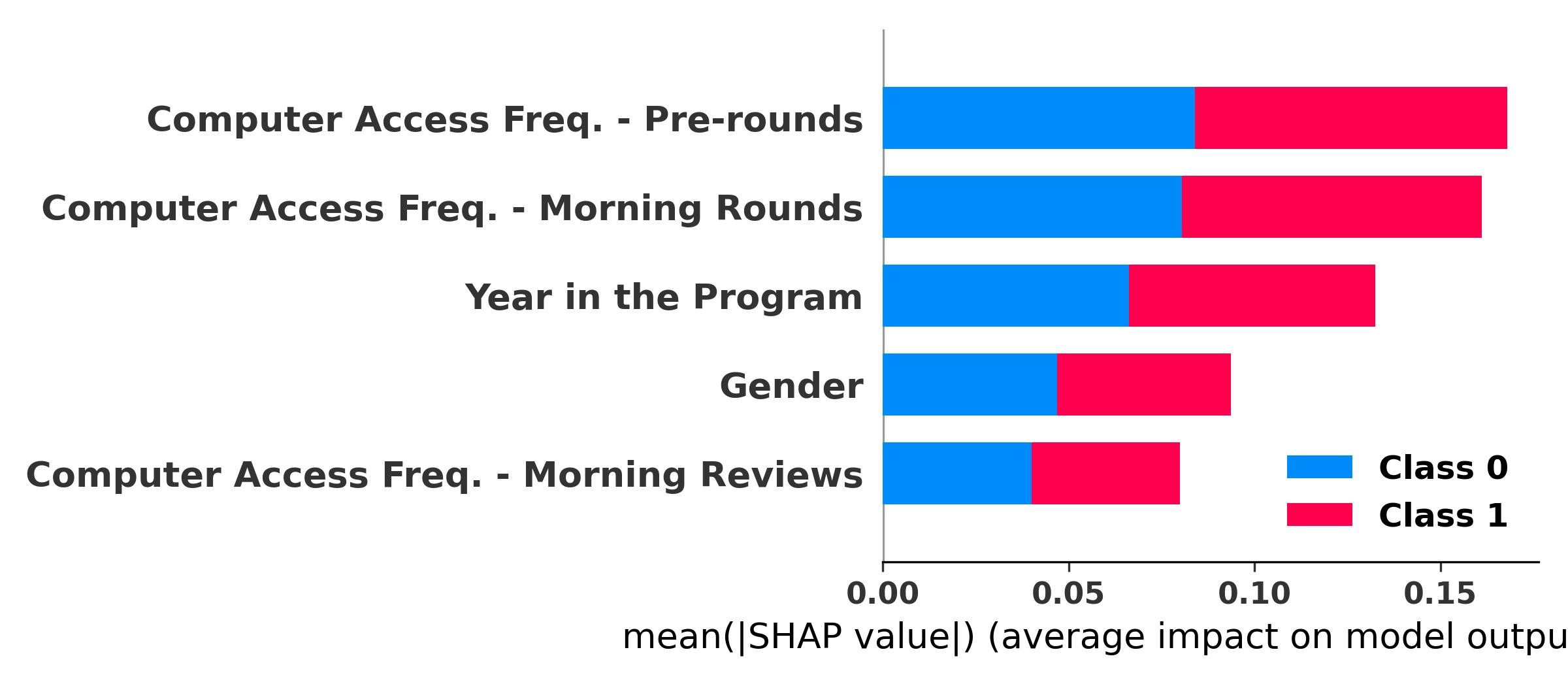}};

        \node at (0.2,4.9) {\footnotesize \textbf{End-of-day Stress - Rest-activity}};
        \node[draw=none,fill=none] at (0,3.3){\includegraphics[width=0.76\linewidth]{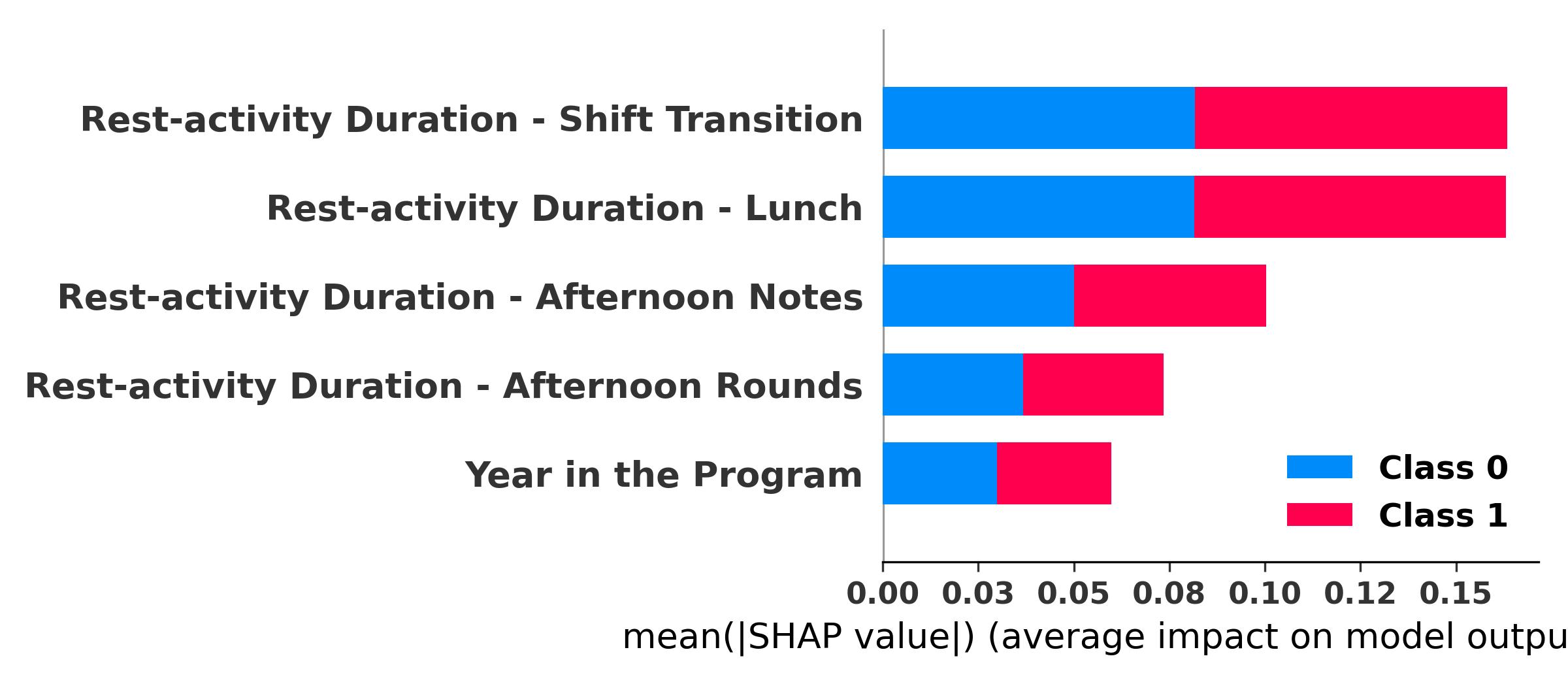}};

        \node at (0.2,1.62) {\footnotesize \textbf{Job Satisfaction - Multimodal}};
        \node[draw=none,fill=none] at (0,0){\includegraphics[width=0.76\linewidth]{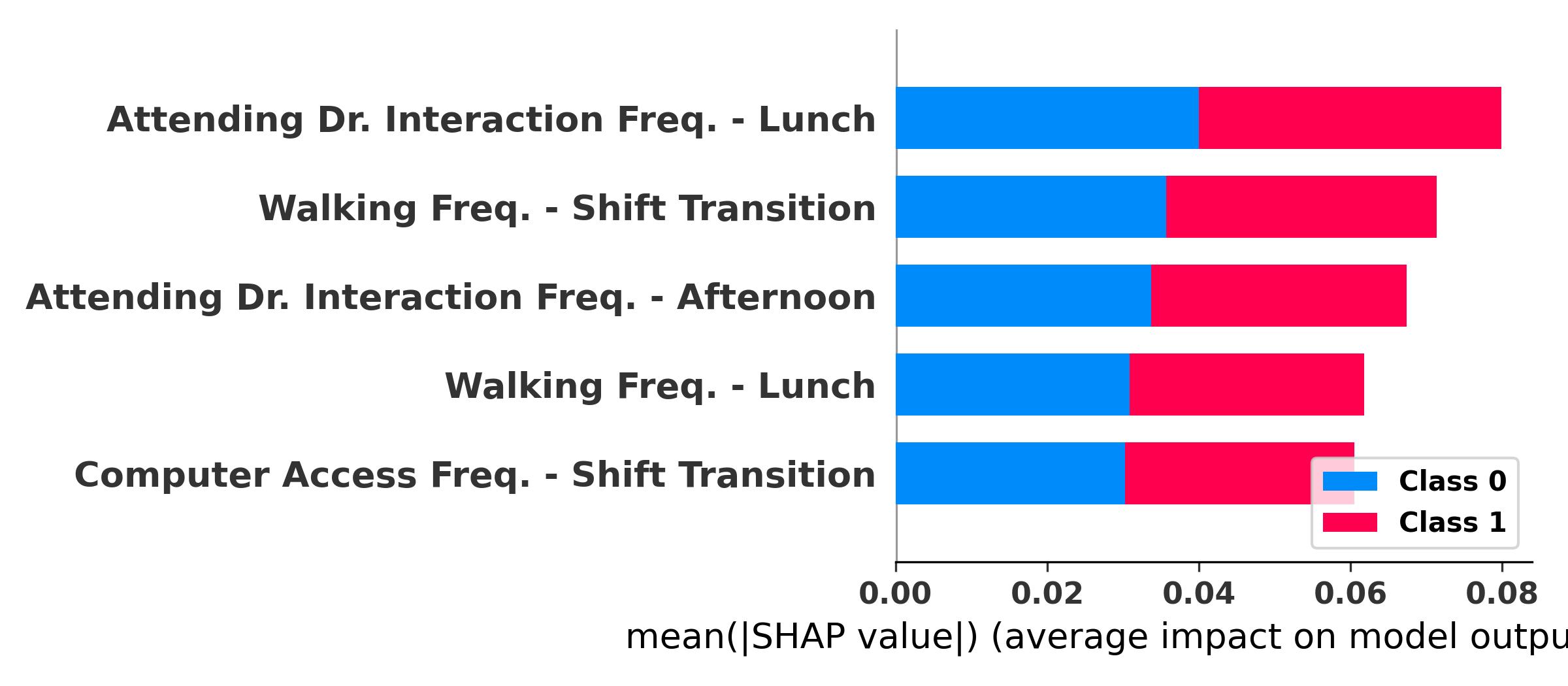}};

    \end{tikzpicture}
    \vspace{-3mm}
    \caption{Feature importance in predicting self-report EMAs.}
    \label{fig:feature_importance}
    \vspace{-3.5mm}
} \end{figure}

\subsection{Machine Learning Results}
\noindent \textbf{Prediction Results} Table~\ref{tab:ml_result} compares the best prediction results (on the macro-F1 score) of the Random Forest classifier among behavioral feature sets we derived in this work. The results indicate that our best prediction results on job satisfaction scores are achieved using multimodal features combining movement, activity, computer access, and doctor interactions. However, the uni-modal feature set yields the best results in predicting midday and end-of-day stress. Particularly, computer access and rest-activity features are indicative of midday and end-of-day stress, respectively. 

\vspace{0.5mm}
\noindent \textbf{Prediction Interpretation} To better understand our model, we report the top 5 most important features in our Random Forest model in Figure~\ref{fig:feature_importance} using the tree explainer toolkit from Shap \cite{lundberg2020local}. From the feature important plot, we can identify that computer access frequency is the most important feature to indicate midday stress, corresponding to our findings from the correlation plot in Figure~\ref{fig:corr_computer}. Moreover, time spent in the rest-activity in the shift transition and lunch are important features in predicting end-of-day stress. Finally, we can observe that interaction with attending doctors is an important factor in determining daily job satisfaction. These results show the effectiveness of our behavioral features in predicting EMAs.

\section{Conclusion}

In this work, we study the behavioral patterns of residents using longitudinal wearable recordings during workshifts. The wearable sensors in this data analysis include a Fitbit wristband, a customized social activity badge, and Bluetooth hubs installed throughout the hospital units. Our data were collected from residents with different training stages and experiences in a complex hospital environment over a 3-week period. Our analysis shows promises of using wearable technologies to study the behavioral patterns of residents, revealing the following unique findings:

\begin{itemize}[leftmargin=*]
    \item Self-reported measure comparisons indicate that residents experience greater negative psychological impacts (e.g., anxiety) than other healthcare providers, such as nurses.
    
    \item Compared to resident interns, the junior and senior residents engage in more walking movement and interactions with the attending doctor on a work shift.

    \vspace{0.5mm}

    \item The workflow level analysis indicates that junior and senior residents have more rest-activity in the morning, more frequent computer access in the afternoon rounds, and more fellow-doctor interaction in the morning than interns.
    
    \vspace{0.5mm}
    
    \item The correlation results reveal that interactions with attending doctors positively correlate with stress and burnout among junior and senior residents, while computer access positively correlates with stress among intern residents.

    \vspace{0.5mm}
    
    \item Our machine learning experiments in predicting EMAs show the effectiveness of different sets of features in predicting stress and job satisfaction. We identify that computer access and rest-activity are midday and end-of-day stress predictors, respectively. Interaction with attending doctors during lunch is important in predicting job satisfaction.

\end{itemize}

\section{Acknowledgement}
The research is based upon work supported by the Office of the Director of National Intelligence(ODNI), Intelligence Advanced Research Projects Activity(IARPA), via IARPA Contract No $2017$ - $17042800005$. The views and conclusions contained herein are those of the authors and should not be interpreted as necessarily representing official policies or endorsements, either expressed or implied, of ODNI, IARPA, or the U.S. Government. The U.S. Government is authorized to reproduce and distribute reprints for Governmental purposes notwithstanding any copyright annotation thereon.

\bibliographystyle{IEEEtranS}
\bibliography{ref}

\end{document}